\documentstyle[12pt]{article}	
\begin{document}

\title{{\bf Nothing but Relativity}}

\author{\bf Palash B. Pal\\ 
\normalsize Saha Institute of Nuclear Physics,
1/AF Bidhan-Nagar, Calcutta 700064, INDIA}

\date{}
\maketitle 

\begin{abstract} \normalsize\noindent
We deduce the most general space-time transformation laws consistent
with the principle of relativity. Thus, our result contains the
results of both Galilean and Einsteinian relativity. The velocity
addition law comes as a bi-product of this analysis.  We also argue
why Galilean and Einsteinian versions are the only possible
embodiments of the principle of relativity.
\end{abstract}
\bigskip\bigskip

Historically, Einstein's special theory of relativity was motivated 
by considerations of properties of light. Even today, textbooks and
other expositions of the special theory rely heavily on gedanken 
experiments involving light. The Lorentz transformation equations,
the formula for relativistic addition of velocities, and other
important formulas of the special theory are derived using light
signals.

There are two assumptions, or axioms, underlying the special theory
of relativity. One is the principle of relativity, which
asserts that physical laws appear the same to any inertial observer.
The other, which marks the difference of Einstein's theory with
the earlier Galilean theory of relativity, is the assertion of the
constancy of the speed of light in the vacuum.

An interesting question to ask, therfore, is the following. Suppose
one takes the principle of relativity, but does not take the second
axiom of Einstein. One would then obtain the most general formulas
equations which are consistent with the principle of relativity. Such
formulas would contain both Galilean and Einsteinian results. This
question has been asked before in the literature \cite{Wei65, Mit66,
LeKa75, Lev76, Rin77, Sri81, Mer84, Sch84, Sch85, Sin86, Sen94}, and
the authors have derived the relativistic velocity addition law in
some cases, the space-time transformation equations in some
other. Here, we present an approach to the same problem which is
somewhat different, and at the end, both the space-time
transformations and the velocity addition law come out from the same
exercise.

Let us consider two inertial frames $S$ and $S'$, where the second one
moves with a speed $v$, along the $x$-axis, with respect to the first one.
The co-ordinates and time in the $S$-frame will be denoted by $x$ and $t$,
and in the frame $S'$, they will be denoted with a prime. The space-time
transformation equations have the form
\begin{eqnarray}
x' &=& X(x,t,v) \,,
\label{x'}\\*
t' &=& T(x,t,v)	\,,
\label{t'}
\end{eqnarray}
and our task is to determine these functions. A few properties of these 
functions can readily be observed. First, the principle of relativity
tells us that if we invert these equations, we must obtain the same
functional forms:
\begin{eqnarray}
x &=& X(x',t',-v) \,,
\label{x} \\*
t &=& T(x',t',-v)	\,.
\label{t}
\end{eqnarray}
Notice that here the third argument of the functions is $-v$, since that
is the velocity of the frame $S$ with respect to $S'$. Using Eqs.\
(\ref{x'}) and (\ref{t'}) now, we can rewrite Eqs.\ (\ref{x}) and
(\ref{t}) as:
\begin{eqnarray}
x &=& X \left(X(x,t,v), \, T(x,t,v), \, -v \right) \,,
\label{Ximpl}\\*
t &=& T \left(X(x,t,v), \, T(x,t,v), \, -v \right) \,,
\label{Timpl}
\end{eqnarray}
which are implicit constraints on the forms of the functions. Moreover,
isotropy of space demands that we could take the $x$-axis in the reverse
direction as well. In this case, both $x$ and $v$ change sign, and so does
$x'$. In other words,
\begin{eqnarray}
X(-x,t,-v) &=& -X(x,t,v) \,, 
\label{X->-X} \\*
T(-x,t,-v) &=& \phantom{-} T(x,t,v)	\,.
\label{T->T}
\end{eqnarray}

We now invoke the homogeneity of space and time. Suppose there
is a rod placed along the $x$-axis such that its ends are at
points $x_1$ and $x_2$ in the frame $S$, with $x_2>x_1$. In the
frame $S'$, the ends will be at the points $X(x_1,t,v)$ and
$X(x_2,t,v)$, so that the length would be 
\begin{eqnarray} 
l' = X(x_2,t,v) - X(x_1,t,v) \,.
\label{l'}
\end{eqnarray}
Suppose we now displace the rod such that its end which used to be at
$x_1$ is now at the point $x_1+h$. Its length in the frame $S$
should not be affected by its position on the $x$-axis by virtue
of the principle of homogeneity of space, so that its other end
should now be at the point $x_2+h$. In the frame $S'$, 
its ends will be at the points $X(x_2+h,t,v)$ and
$X(x_1+h,t,v)$. However, homogeneity of space implies that
the length of the rod should not be affected in the frame $S'$
as well, so that
\begin{eqnarray}
l' = X(x_2+h,t,v) - X(x_1+h,t,v) \,.
\label{l'h}
\end{eqnarray}
Using Eqs.\ (\ref{l'}) and (\ref{l'h}), we obtain
\begin{eqnarray}
X(x_2+h,t,v) - X(x_2,t,v) = X(x_1+h,t,v) - X(x_1,t,v) \,.
\end{eqnarray}
Dividing both sides by $h$ and taking the limit $h\to 0$, we
obtain 
\begin{eqnarray}
\left. {\partial X \over \partial x} \right|_{x_2} =
\left. {\partial X \over \partial x} \right|_{x_1} \,.
\end{eqnarray}
Since the points $x_2$ and $x_1$ are completely arbitrary, this
implies that the partial derivative 
$\partial X/\partial x$ is constant, independent of the point
$x$. Thus, the function $X(x,t,v)$ must be a linear function of
$x$. One can similarly argue, invoking the homogeneity of time
as well, that both $X(x,t,v)$ and $T(x,t,v)$ are linear in the
arguments $x$ and $t$. In that case, making the trivial choice
that the origins of the two frames coincide, i.e., $x=t=0$
implies $x'=t'=0$, we can write
\begin{eqnarray}
X(x,t,v) &=& A_v x + B_v t \,, 
\label{Xlin} \\*
T(x,t,v) &=& C_v x + D_v t \,,
\label{Tlin}
\end{eqnarray}
where the subscript $v$ on the co-efficients $A$, $B$, $C$ and $D$ 
remind us that they are
functions of the relative velocity $v$ only. 
Eqs.\ (\ref{X->-X}) and (\ref{T->T}) then imply that
\begin{eqnarray}
A_{-v} = A_v \,, \quad B_{-v}=- B_v \,, 
\quad C_{-v}= -C_v \,, \quad D_{-v}= D_v \,.
\label{-v}
\end{eqnarray}
In other words, $A$ and $D$ are even functions, while $B$ and $C$ are
odd functions of $v$. Using these properties, we can now use Eqs.\
(\ref{Ximpl}) and (\ref{Timpl}) to obtain the following conditions:
\begin{eqnarray}
A_v^2 - B_v C_v &=& 1 \,, \label{cond1} \\*
B_v (A_v - D_v) &=& 0 \,, \label{cond2} \\*
C_v (A_v - D_v) &=& 0 \,, \label{cond3} \\*
D_v^2 - B_v C_v &=& 1 \label{cond4} \,.
\end{eqnarray}
Unfortunately, these four equations do not solve the four functions
$A$, $B$, $C$ and $D$. The reason is simple. Eqs.\ (\ref{cond2}) and
(\ref{cond3}) indicate two possibilities. Either $B_v=C_v=0$, in
which case the other two equation say that $A_v=D_v=1$, which is just
the trivial solution of identity transformation. While this is
mathematically a valid possiblity, physically it is not acceptable
for arbitrary values of $v$.
Thus we look at the other case, which gives
\begin{eqnarray}
D_v &=& A_v \,,  \label{D=A} \\*
C_v &=& {A_v^2 -1 \over B_v} \,.
\label{C}
\end{eqnarray}
Thus, two of the functions of $v$ introduced in Eqs.\ (\ref{Xlin})
and (\ref{Tlin}) are independent.

In fact, we can reduce the number of independent functions further if
we notice that by our definition, the origin of the frame $S'$ is
moving at a speed $v$ with respect to the origin of $S$, i.e., at
time $t$, it must be at the point $x=vt$. In other words, $x'=0$ when
$x=vt$. This implies
\begin{eqnarray}
B_v = -vA_v \,, 
\end{eqnarray}
so that now we can write down the transformation equation in terms of
just one unknown function $A_v$, this time in a matrix notation:
\begin{eqnarray}
\left(\begin{array}{c} x' \\ t' \end{array}\right) = 
\left(\begin{array}{cc} A_v & -vA_v \\ -{A_v^2-1 \over vA_v} & A_v
\end{array}\right)  
\left(\begin{array}{c} x \\ t \end{array}\right) \,.
\label{matrix}
\end{eqnarray}

So far, the functional form of $A_v$ is unknown, except for the fact
that it is an even function of $v$, and that it must equal unity when
$v=0$. However, we can go further if we now consider a third frame
$S''$ which is moving with a speed $u$ with respect to $S'$. Then
\begin{eqnarray}
\left(\begin{array}{c} x'' \\ t'' \end{array}\right) &=& 
\left(\begin{array}{cc} A_u & -uA_u \\ -{A_u^2-1 \over uA_u} & A_u
\end{array}\right)  
\left(\begin{array}{cc} A_v & -vA_v \\ -{A_v^2-1 \over vA_v} & A_v
\end{array}\right)  
\left(\begin{array}{c} x \\ t \end{array}\right) \nonumber\\
&=& 
\left(\begin{array}{cc} A_u A_v + (A_v^2-1) {uA_u \over vA_v} 
& -(u+v)A_u A_v \\ 
-(A_u^2-1) {A_v\over uA_u} - (A_v^2 -1) {A_u \over v A_v} & A_u A_v +
(A_u^2-1) {vA_v \over uA_u}
\end{array}\right)  
\left(\begin{array}{c} x \\ t \end{array}\right) 
\,.
\label{S"S}
\end{eqnarray}
However, Eq.\ (\ref{D=A}) tells us that the two diagonal elements of 
this matrix should be equal, which implies
\begin{eqnarray}
{A_v^2 -1 \over v^2 A_v^2} = {A_u^2 -1 \over u^2 A_u^2} 
\label{K}\,.
\end{eqnarray}
But the left side of this equation depends only on $v$, while the
right side depends only on $u$. They can be equal only if they are
constants. Denoting this constant by $K$, we obtain
\begin{eqnarray}
A_v = {1 \over \sqrt{1 - Kv^2}} \,.
\label{A}
\end{eqnarray}
Using this form in Eq.\ (\ref{matrix}), we thus obtain that the most
general transformation equations consistent with the principle of
relativity are of the form
\begin{eqnarray}
\left(\begin{array}{c} x' \\ t' \end{array}\right) = 
{1 \over \sqrt{1 - Kv^2}} 
\left(\begin{array}{cc} 1 & -v \\ -Kv & 1 
\end{array}\right)  
\left(\begin{array}{c} x \\ t \end{array}\right) \,.
\label{general}
\end{eqnarray}

Another thing to notice is that the velocity addition law can be
directly deduced from our analysis. For this, let us call the speed
of the frame $S''$ with respect to $S$ by $w$. Then, in Eq.\
(\ref{S"S}), the diagonal terms of the matrix must be $A_w$:
\begin{eqnarray}
A_w &=& A_u A_v + (A_v^2-1) {uA_u \over vA_v}  \nonumber\\ 
&=& A_u A_v (1+Kuv) \,,
\label{Aw}
\end{eqnarray}
using in the last step the definition of $K$ which follows from
Eq.\ (\ref{K}). 
Given the form of the function $A$ from Eq.\ (\ref{A}), it is
now easy to deduce that
\begin{eqnarray}
w = {u+v \over 1+Kuv} \,,
\label{addn}
\end{eqnarray}
which is the velocity addition law.

Specific theories of relativity, of course, have to make extra
assumptions in order to determine the value of $K$. In the case
of Galilean relativity, this extra assumption shows up in the
form of the universality of time, which means $t'=t$ for any
$v$. Obviously, this requires $K=0$. The extra assumption for
Einstein's theory of relativity is the constancy of the speed of
light in vacuum. From Eq.\ (\ref{addn}), it is easy to see that
$K^{-1/2}$ is an invariant speed, independent of the frame of
reference. Thus, $K=1/c^2>0$ in this case. It is obvious that in
both these cases, we obtain the appropriate transformation laws
from Eq.\ (\ref{general}) and the velocity addition law from
Eq.\ (\ref{addn}).

From this line of reasoning, it seems that there should be
another logical possibility with $K<0$. Actually, this option is
not self-consistent. To see this, we first look at Eq.\
(\ref{A}), and note that only the positive square root can be
taken in the expression on the right hand side, because we want
$A_v$ to reduce to unity when $v$ vanishes. Thus, $A_v \geq 0$ for any
$v$. However, if $K$ is negative, i.e., $K=-1/C^2$ for some finite value
of $C$, we can obtain 
$A_w<0$ from Eq.\ (\ref{Aw}) if we choose large enough values of
$u$ and $v$ which satisfy $uv>C^2$. 

One point has to be made here. For the case of Einsteinian relativity as
well, one can reach a contradiction, viz., that $A_v$ becomes imaginary if
$v>c$. But such large speeds are unreachable in Einsteinian relativity due
to the structure of the addition law of Eq.\ (\ref{addn}), which shows
that one cannot obtain $w>c$ if both $u$ and $v$ are less than $c$. For
$K=-1/C^2$, this is not the case. One can add two speeds, both less than
$C$, and the result of addition can be larger than $C$. For example,
if the speed of $S'$ is $C/2$ with respect to $S$, and if $S''$ moves with
a speed $C/2$ with respect to $S'$, the speed of $S''$ from the $S$-frame
is $4C/3$. Thus, speeds larger than $C$ cannot be excluded from this
theory, but such speeds raise the possibility of having $A_w<0$ as
outlined above. Hence the inconsistency.

Thus, in effect, we have deduced the most general space-time
transformation law as well as the velocity addition law
consistent with the principle of relativity, and have shown that
Galilean and Einsteinian laws are the only possible ones.
Our method most closely resembles that of Singh \cite{Sin86}, but
there are important differences. In his derivation, Singh used some
properties of the velocity addition law deduced by Mermin
\cite{Mer84}. We have not used them. On the other hand, we have made
direct use of the isotropy of space to deduce the symmetry properties
of the functions $A$, $B$, $C$ and $D$ which have been summarized in
Eq.\ (\ref{-v}) and used them to obtain Eqs.\
(\ref{cond1}-\ref{cond4}). But the most important difference, to our mind,
is that while previous derivations used distinct lines of
reasoning for the  space-time transformation laws and the
velocity addition formula, our argument
gives both  at the same stroke.

\paragraph*{Acknowledgements:} I thank P. Bhattacharjee, B. P.
Das, J. Samuel, D. Sen and S. Sinha for patiently listening to my
arguments, commenting on them, and suggesting some references which I
could have missed otherwise.



\begin{thebibliography}{WW}
\bibitem{Wei65} R. Weinstock: {\sl New approach to special relativity},
Am. J. Phys. 33 (1965) 540---645.

\bibitem{Mit66} V. Mitavalsk\'y: {\sl Special relativity without the
postulate of constancy of light}, Am. J. Phys. 34 (1966) 825.

\bibitem{LeKa75} A. R. Lee, T.~M. Kalotas: {\sl Lorentz
transformations from the first postulate}, Am. J. Phys. 43
(1975) 434---437.

\bibitem{Lev76} J.-M. L\'evy-Leblond: {\sl One more derivation
of the Lorentz transformation}, Am. J. Phys. 44 (1976) 271---277.

\bibitem{Rin77} W. Rindler: {\sl Essential Relativity}
(Springer-Verlag, 2nd edition, 1977). See \S 2.17.

\bibitem{Sri81} A. M. Srivastava: {\sl Invariant speed in special
relativity}, Am. J. Phys. 49 (1981) 504---505.

\bibitem{Mer84} N. D. Mermin: {\sl  Relativity without light}, Am.
J. Phys. 52 (1984) 119---124.

\bibitem{Sch84} H.~M. Schwartz: {\sl Deduction of the general
Lorentz transformations from a set of necessary assumptions},
Am. J. Phys. 52 (1984) 346---350.

\bibitem{Sch85} H.~M. Schwartz: {\sl A simple new approach to
the deduction of the Lorentz transformations},
Am. J. Phys. 53 (1985) 1007---1008.

\bibitem{Sin86} S. Singh: {\sl Lorentz transformations in Mermin's
relativity without light}, Am. J. Phys. 54 (1986) 183-184.

\bibitem{Sen94} A. Sen: {\sl How Galileo could have derived the special
theory of relativity}, Am. J. Phys. 62 (1994) 157-162.

\end{thebibliography}
\end{document}